# ICT, Community Memory and Technological Appropriation

Rodrigo Torréns
Corporación Parque Tecnológico de Mérida,
Venezuela

Luis A. Núñez
Universidad de Los Andes, Mérida, Venezuela

Raisa Urribarri
Universidad de Los Andes, Trujillo, Venezuela

The core mission of universities and higher education institutions is to make public the results of their work and to preserve the collective memory of the institution. This includes the effective use of information and communication technologies (ICT) to systematically compile academic and research achievements as well as disseminate effectively this accumulated knowledge to society at large. Current efforts in Latin America and Venezuela in particular, are limited but provide some valuable insights to pave the road to this important goal. The institutional repository of Universidad de Los Andes (ULA) in Venezuela (www.saber.ula.ve) is such an example of ICT usage to store, manage and disseminate digital material produced by our University. In this paper we elaborate on the overall process of promoting a culture of content creation, publishing and preservation within ULA.

## Introduction and General Considerations

Due to the development and dissemination of Information and Communication Technology (ICT), there are greater opportunities to publish and access research results and intellectual production at university institutions. The academic use of these technologies, and in particular Institutional Repositories (IIRR), is essential to reach goals and milestones related to the preservation and publication of scientific and humanitarian heritage produced in education centers, which also contributes to the preservation of their institutional memory.

Without a doubt, university communities made up of teachers, researchers and students are a permanent source of knowledge production. This is a system in which, as Peset et al (2005) point out, although the actors produce the information, they then have to pay in order to access it, acquiring publications through personal or institutional subscriptions. To correct this situation, traditional dissemination and publication models must change. In the same way, the quality, accessibility and preservation in time of intellectual heritage must be guaranteed in order to reach the goals set by the Budapest Open Access Initiative (BOAI)[72] and the Berlin Declaration on free access to humanistic, artistic and scientific knowledge[73]. One of the main ideas behind these initiatives is that free and open access to knowledge generates in turn more knowledge and benefits for humanity; any kind of control or restrictions on this knowledge would be an obstacle for the advancement of the sciences (Guedon, 2002).

## ICT and digital preservation

Current research results are mostly published in digital format. Electronic publishing has helped to change the way in which the actors involved in the editing process relate and work (universities, researchers, publishing houses and libraries (Steenbakkers, 2003). According the digital encyclopedia Wikipedia[74,] digital preservation can be considered as the group of processes and activities that ensure the continuous long-term access to existing information and scientific registries and to cultural heritage in electronic formats

It could be said that thanks to digital technologies the preservation of knowledge is an easier process, but it is not so. Although we can still read on paper materials written hundreds of years ago, digital information created only a decade ago is in serious danger of being lost. This is why some countries have decided to save and preserve valuable information. One of these projects is the Digital Longevity Project, undertaken by the National Archive of the Netherlands[75].

---

[72] Budapest Open Access Initiative: http://www.soros.org/openaccess

[73] Original text for the Berlin Declaration http://www.zim.mpg.de/openaccess-berlin/berlindeclaration.html

[74] http://en.wikipedia.org/wiki/Digital_preservation

[75] Digital Longevity Project:

## Institutional Repositories as tools for the preservation of institutional memory

Institutional Repositories are digital collections that capture, preserve and disseminate the intellectual production of university communities (Crow, 2002). A university institutional repository (IR) is "a group of services that a university offers to the members of its community to create and disseminate digital material created by the institution and the members of its community. At the same time, it is an institutional commitment for the safeguard of these digital materials, including their long-term preservation, organization, access and distribution" (Lynch, 2003).

The IR is a digital archive of the intellectual production generated by professors, researchers, employees and students of an institution which is accessible to the end-users, both inside and outside the institution, with little or no obstacles to access it. According to Crow (2002), one of the main characteristics of an IR is that it is defined and supported by the institution. Also, it has academic and research purposes. From the information point of view, the IR is cumulative, permanent, open and interoperating. In other words, an IR is a place of easy and free access that hosts "treasures" (Drake, 2004) previously hidden, where both experts and amateurs can go to searching for answers and benefit from the collective knowledge it holds.

Collective memory is a term coined by the French philosopher and sociologist Maurice Halbwachs (Rheims, March 1877–Buchenwald, March 1945)[76] and, contrary to individual memory, it is shared, transferred and built by a group, institution, community or society as a whole. The preservation of collective institutional memory depends on several elements: the conscience that members of the institution have of the importance of this task, in the first place, but also of the existing abilities to preserve data and the information produced in the institution, as well as to analyze it and process it.

IIRR play an important role in the dynamics of preserving and disseminating the institutional or collective memory. Also, a group of them may be considered jointly as part of the collective memory of a whole country or culture. As an example, we can mention the "Digital Academics Repositories" (DARE)[77] project in the Netherlands, in which

---

http://www.digitaleduurzaamheid.nl/index.cfm?paginakeuze=286&categorie=6

[76] Collective Memory and Time. Maurice Halbwachs. http://www.uned.es/ca-bergara/ppropias/vhuici/mc.htm

[77] Proyecto DARE: http://www.surf.nl/en/themas/index2.php?oid=7

the IIRR are the base of the common infrastructure of electronic publication that preserves and spreads the intellectual production of all the universities in the country.   DAREnet[78,] on the other hand, offers a common access to the research results that are produced in different institutions.

This contribution aims to systematize and share the experience of over a decade in the development and operation of the Institutional Repository (IR) of the University of Los Andes (ULA, Venezuela). We present a panoramic vision of the processes that we have followed for the promotion of the culture to capture, preserve and disseminate contents through an IR. We will show some quantitative results related to the operation of the repository, and we will present some final general ideas on the hardships and challenges that came with the development of the IR at the ULA.

## The SABER-ULA IR: a methodological approach

The ULA has around 40 thousand students and five thousand professors and researchers.  It is made up of eleven schools in which undergraduate and graduate courses are taught, related to different knowledge areas (engineering, technology, health sciences, social sciences, and so on).  It has around 200 research groups and offers approximately 160 postgraduate programs.  It has three branches in three states in western Venezuela (Táchira, Mérida and Trujillo.)  The main campus is in Mérida (capital of the state with the same name), a small city in the Venezuelan Andes with an approximate population of half a million people and is characterized by having inhabitants that uses ICT in higher proportion than the rest of the country, as pointed out by the Human Development Report by the UNDP for the year 2002 (UNDP, 2002).

As mentioned by Dávila et al (2006-1), the ULA has had a leading role in building a welcoming environment for technological innovation.  In this sense, the ULA, along with the Mérida Technological Park Corporation[79] (CPTM), has began to build, since the year 2000, a university IR aiming to save and spread out the production and university intellectual heritage. The project received the name of "SABER-ULA, the Intellectual Heritage of the ULA on the Internet."[80]

Taking the ideas from the international movement for Open Access to knowledge proposed at the BOAI, the Berlin Declaration, as well as our own experiences, as our foundations; we believe that the process of appropriation of ideas and tools that promote the free dissemination of knowledge produced in our institutions is related to the following elements:

1. Creation of a sustainable infrastructure to handle information
2. Identification of the individuals and/or communities that produce information and the incorporation of these in the process of publication, dissemination and preservation of digital content.
3. Design and application of an appropriate methodology for training and orientation in handling the tools by the actors involved.
4. Generation of policies for handling information and incentives for the producers of said information.
5. Promote the use of the IR contents.

---

[78] DARENet: http://www.darenet.nl/
[79] CPTM: http://www.cptm.ula.ve
[80] Web portal of IR SABER-ULA: http://www.saber.ula.ve

With our experience as the example, each one of these elements is described as follows:

## Creation of a sustainable infrastructure to handle information

The first step is to create a specialized work unit, made up of experts in handling the information. At the beginning, this unit will be in charge of generating the basic technological conditions to fulfill the stated objectives. This means selecting, testing and adapting the available technological tools; plan and organize the models and procedures for offering the services; and, prepare presentations and talks for the members of the academic community, among other activities.

An important task is to develop and offer, from the beginning, services with added value (search and capture of information, meet international standards of interoperability, etc) for the contents of the IR. This unit also has the tasks of offering user help; maintain the operability of the infrastructure and the services it offers, as well as safeguarding the stored digital objects, one of the critical tasks in an IR. A key element is the permanent awareness of technological changes that in this day and age happen very quickly

The activities mentioned previously require organizational and operative structures dedicated exclusively to maintaining these services 24 hours a day, 7 days a week (Dávila, J. et al., 2006-1). In the case of the IR SABER-ULA, the development and maintenance of the services is guaranteed by personnel hired specifically for this task, through the management of the CPTM, an organization created by the ULA to operate and manage its tele-information services.

## The process of publication, dissemination and preservation of digital content

The second crucial step for the development of an IR, in parallel to the previous point, is to identify key and proactive communities in the production of information and its electronic management.  These communities can be academic departments, research units, administrative dependencies, laboratories, etc. (Barton and Waters, 2004).  The academic and research community is a natural channel and easily permeated with use and adoption of ICT (Dávila et al, 2006-2); but the ones chosen should be those that offer less resistance to experimentation and are aware of the advantages that the widespread use and knowledge of ICT can have.

In the instance of the IR at the ULA, some successful strategies have included contact with the researchers and more productive groups, according to the national and institutional evaluation and promotion systems[81]; awareness-raising of those responsible for research units and academic departments; offering electronic publishing services to the editors of arbitrated scientific journals produced by members of the institution; maintaining a flowing communication relation with the communities that provide content, and most importantly, the creation of trust relations between the communities and the team in charge of the repository, based on the professional support of excellence (Barton y Waters, 2004).

The criteria for selecting key communities can also be related to identifying leaders in each faculty or campus, or by incorporating groups and people who are convinced of the benefits generated by an IR.

## Design and application of an appropriate methodology for training and orientation in handling the tools by the actors involved

It is important to offer training services to the members of the academic community that have or will have content collections to feed the IR.  Researchers can be informed and trained individually in the use of the necessary tools, or a responsible member can be trained in each community (research unit, department, etc.) to be in charge of adding the content they generate to the IR.  This will depend on the publishing model chosen for the IR and its services.  Personnel whot will add metadata to the contents and offer service support must also be trained, as well as the organizing managers and technicians involved.  It is important to update the IR personnel with emerging technologies, new platforms and programming languages, which will be a good investment at the time when changes are made to the technological systems that support the repository.

---

[81] Research Stimulus Program (PEI) and Direct Support for Groups Program (ADG) at every institution. In the ULA:
http://www.ula.ve/cdcht/prog_investigacion/prom_est.php;        and        Researcher Stimulus  Program (PPI) at the national level in Venezuela: http://www.ppi.org.ve/

## Generation of policies for handling information and incentives for the producers of said information

Although many institutions have formal decrees, resolutions or at least recommendations related to the deposit in an IR of the contents generated by the institution, it is vital that the authorities recognize that the maintenance of their IR is a commitment and a long-term institutional task that must  be a part of  the institutional policies.  In addition, financing policies for the operation of the IR, as well as those aimed to promote the creation of content are necessary.

In the last follow-up meeting for the Berlin Declaration (Berlín 3 Open Access)[82] it was agreed that all institutions must comply with the following: 1. Require a copy of all articles published in an open access repository from the researchers, and, 2.  Encourage researchers to publish their articles in existing open access journals.  Also, it was requested that the institutions register e-prints of their commitment with the institution[83] and describe their policies[84].

It is important to develop and incorporate incentive policies and acknowledgements to the information producers that publish content in repositories and open access journals.  At the same time, guidelines and clarification must be given for the issues related to authors' rights and copies of the content generated by the institution.

---

[82] Berlin 3 Open Access Outcomes:
 http://www.eprints.org/events/berlin3/outcomes.html
[83] To sign the commitment and register the institution's policy:
 http://www.eprints.org/signup/sign.php.
[84] List of institutions and policies:
http://www.eprints.org/signup/fulllist.php.

## Promote the use of the IR contents

Once the IR is built, it is then critical to communicate the benefits that it offers to the university community (Barton and Waters, 2004). This can be achieved in two ways, from top to bottom or bottom to top. The first implies forming leaders and institution authorities, deans, etc; developing pilot communities for demonstration purposes before the rest of the institution. The second means informing the content producers (researchers and research groups, professors, technical and administrative personnel, librarians, etc) through direct presentations to the members of the university community, promotion through institutional and local press, brochures and posters, and using publicity mediums inside and outside the university.

## Development and consolidation of the IR

The development of the SABER-ULA IR (2000-2006) as a preservation and dissemination tool for the intellectual production of the members of the university community at the University of Los Andes[85], has occurred in three well-defined phases, each one lasting two years, of infrastructure building, consolidation of service and acknowledgment on behalf of the users.

### Phase I: Construction of the basic working infrastructure

In the first phase (2000-2002), different information and communication sessions were held regarding the IR services for researchers and research units. Thanks to the contribution of other work units related to ICT that had organizational and financial recognition by his institution, the technological, organizational and administrative infrastructure was created to take on the tasks of the IR.

Due to the lack of official institutional support, among other reasons, this approach did not produce the expected results, which is when the services of the IR were offered to the editors if academic, research and informative journals. This way, more content was captured, as the request came directly from editors who turned in between 8 and 10 full articles with each issue of the journal that was to be published. In this phase, the first 10 electronic journals were created. Also at this time was created a repository of academic events[86] as a result of a request on behalf of members of the university community that had no way individually or institutionally to publish and disseminate their regular activities.

---

[85] ULA Web portal: http://www.ula.ve

## Phase II: Consolidation of services and new demands

In the second phase (2002-2004) the services of electronic publishing were consolidated and new demands, specific to the university community, arose; this led to the specialization of tasks among those providing the services and the definition of specific procedures for each type of content.

The content creators, especially through the editors of scientific journals, began to demand new services and swiftness in the publishing process. At this time, some editors began using electronic publishing as a substitute for traditional publishing, due to the economic and organizational problems that usually hold back paper publications, threatening the periodicity of some of the journals.

In this phase begins the true adoption of SABER-ULA as a tool that covers different objectives both for the authors of the content as for the institution. Although one could say that this adoption is still informal, as the creators of the content still do not have any kind of recognition and/or promotion for publishing their production electronically.

In such a phase, services without a lot of initial acceptance, such as the maintenance of the database of researchers and research units, begin to generate use from different dependencies within the university. Management tools for the repository are added in a test phase to ensure interoperability of the system with other providers; different events are held related to digital libraries and work is done on generating models to publish thesis in electronic formats.

## Phase III: Recognition of the IR by the users

Between 2004 and 2006, a regular volume was in the processing of content (journal articles, pre-prints, event references, etc.). During the first trimester of this year an average of 500 registries a month were processed. The number of electronic journals reaches 40 and eight thousand registries were published in the IR. The users began to recognize the value of the information held by the IR. Historians from the institution requested use of the registry to build a memory of the events

---

[86] ULA Events portal: http://www.saber.ula.ve/eventos

that took place in the University.

The ULA reached important visibility of its contents on the Internet thanks to the quantity and quality of the IR[87]; however, there was still not a full institutional recognition that could lead to full financing for supporting services. At the end of the first trimester of 2006, the ULA officially declared its commitment to adhere and sign the Berlin Declaration, which meant a great step forward in the understanding of the importance of the ideas held by the movement and the initiatives for open access to information (OAI), in which IIRR play an important role.

## Some significant numbers

### Queries

Since its creation in the year 2000 until March 2006, more than 8 million of searches on documents and information registries have carried out in the IR of the ULA, SABER-ULA. In the last two years (2005-March 2006), as can be seen in the following chart (Figure 1), the increase in the amount of queries has been notable: only in the first three months of the year 2006 the number was above the total for the whole year 2004.

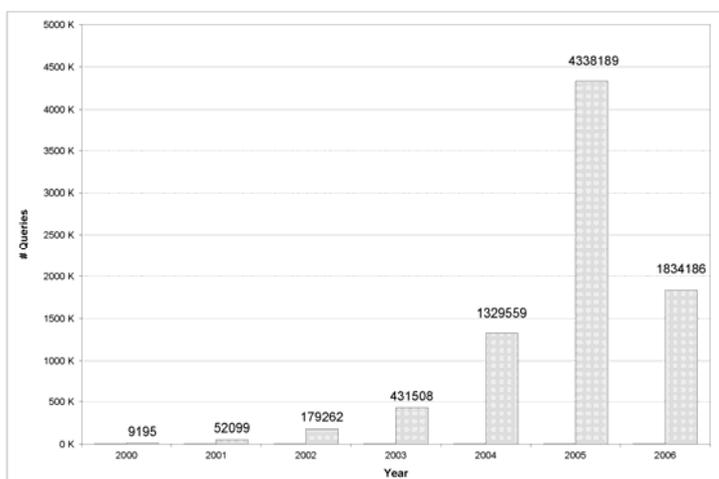

**Figure 1: Annual queries at the Institutional Repository SABER-ULA
(up to March 31, 2006)**

---

[87] See University Visibility Ranking at:
http://www.webometrics.info/top100_continent.asp-cont=latin_america.htm

## Registries and publications

The next figure (Figure 2) represents how the content of the repository has increased substantially year to year since it began offering services. This is a sign of the appropriation and acceptance that the electronic publishing services have had, mainly among the journal editors of the institution. This coincides with the international tendencies reported by Swan and Sheridan (2005). In their annual study on the adoption of Open Access they point out that auto-archiving the use of institutional repositories has increased 60% between 2004 and 2005.

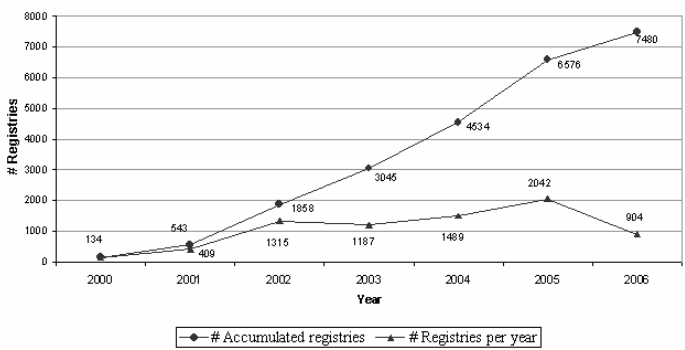

**Figure 2: Number of information registries in the Institutional Repository SABER-ULA (up to March 31, 2006)**

Around 50% of the IR of the ULA follows the "golden path" (Suber, 2005) established in the open access initiatives and the Berlin Declaration; wich means that this important percentage of the IR contents come from electronic university journals.

# Conclusions and future work

According to Peset et al (Peset, F. et al., 2005), the changes that Internet has brought to the communication model reside in the possibility of offering visibility to the scientific production of an institution or a country in ways that were unthought of until recently. The IIRR are one of the main tools to facilitate that change and their appropriation, on behalf of the communities of authors and users of the information, is generating an interesting dynamic of creation, preservation and use of knowledge that spreads to the rest of society.

After six years of development at the IR SABER ULA, today we can say that there is an acknowledgment and institutional recognition of free-access electronic publishing, and that the adoption of ICT has created a demand for new services and requests for improvements of the tools related to electronic publishing. An indicator of this is that close to 90% of the academic journals (at ULA) are being published in electronic format in the IR.

However, although the perceived resistance to the dissemination of the produced information has decreased, there are still some obstacles, among which we can name the following:

- **The lack of incentives for electronic publishing**, which makes it difficult to incorporate authors and communities as collaborators and receptors of the services offered by the repository. To try and remedy this situation, the ULA is recognizing through symbolic prizes and academic events the authors and publications with the highest number of queries through the IR[88].

- **Limited awareness of the need for preservation.** From the beginning, the work team of the repository has constantly contributed to the recovery of valuable digital archives with valuable content to which the author originally did not give the importance to preserve, as the content had already been published on paper (in a journal, a book, etc). With time, this problem decreases, but it still persists, which makes us see that the paper culture is still deeply rooted among the information producers.

- **The lack of training for the appropriate use of ICT**. The two previous problems are related to the deficient culture for the appropriate use of ICT on behalf of those who generate knowledge. Although we have no way to measure this in quantity, we perceive that this situation has decreased progressively at the same time that formal and informal training is offered to the content creators and those involved in the use of tools and digitalization techniques, file formats, creation of digital content, etc.

---

[88] See a list of acknowledgements at: http://www.saber.ula.ve/estadisticas

- **Limited usability of computer technology tools**. The next obstacle in the appropriation process is the fact that the tools used to manage the repository still do not offer the grade of *usability*[89] and *personalization*[90] that would significantly ease the use of these on the users end. Many of the publication processes still depend on the personnel at the repository, which gives the added value to the data sent by the authors of the content.

- **Lack of credibility for the contents available on the Internet**. Although some researchers say they have reservations and distrust for the contents available on the Internet, and thus, don't have an interest in publishing under this modality; they also express fear that their work may be plagiarized or used without the credit for the original source. These reserves decrease with time, but still exist.

- **Irregular and changing political and institutional support**. Important support has been received from the university community and its leaders and authorities, but some sectors of the institution still see the IR as a threat or as unnecessary with the characteristics it currently has. Some have serious objections or doubts over the organizational structure behind the repository and the cost associated with its operation. Others disagree with technical aspects or with the service models implemented.

Currently, activities related to the update of the technological platform that supports the Institutional Repository of the ULA are taking place. There is also work being done, along with the responsible authorities and dependencies, to create and adopt formal policies within the University to promote, or make compulsory, the free dissemination of intellectual production of the institution through IIRR; as many institutions around the world are doing in order to comply with the recommendations from the Berlin Declaration; this will help, in the near future, to overcome some of the obstacles mentioned previously.

The establishment of interoperable networks made up of repositories along Latin America will increase the impact of the content produced in the region and will give it a visibility and use until recently difficult to envision.  We are working on proposals for the development of this kind of initiatives in other institutions in Venezuela and Latin America.

---

[89] See definition of "usability" at: http://es.wikipedia.org/wiki/Usabilidad
[90] See definition of "personalization" at:
http://en.wikipedia.org/wiki/Personalization